\documentclass[12pt]{iopart}
\usepackage{graphicx}
\usepackage{bm}
\usepackage{citesort}
\usepackage{iopams}

\newcommand{\PRB}{\PR B }

\newcommand{\PRE}{\PR E }
\newcommand{\PLA}{{\it Phys. Lett.} A }

\newcommand{\SPJ}{{\it Sov. Phys. - JETP} }
\newcommand{\UFN}{{\it Usp. Fiz. Nauk} }
\newcommand{\PU}{{\it Phys.-Uspekhi} }
\newcommand{\ZETF}{{\it Zh. Eksp. Teor. Fiz.} }
\newcommand{\ZPB}{\ZP B }

\begin{document}
\title[Interplay between boundary conditions and cubicity of Ginzburg-Landau equation]{Influence of the interplay between de Gennes boundary conditions and cubicity of Ginzburg-Landau equation on the properties of superconducting films}
\author{O Olendski} \address{King Abdullah Institute for Nanotechnology, King Saud University, P.O. Box 2454, Riyadh 11451, Saudi Arabia}
\ead{oolendski@ksu.edu.sa}

\begin{abstract}
Exact solutions of the Ginzburg-Landau (GL) equation for the straight film subjected at its edges to the Robin-type boundary conditions characterized by the extrapolation length $\Lambda$ are analyzed with the primary emphasis on the interaction between the coefficient $\beta$ of the cubic GL term and the de Gennes distance $\Lambda$ and its influence on the temperature $T$ of the strip.  Very substantial role is played also by the carrier density $n_s$ that naturally emerges as an integration constant of the GL equation. Physical interpretation of the obtained results is based on the $n_s$-dependent effective potential $V_{eff}({\bf r})$ created by the nonlinear term and its influence on the lowest eigenvalue of the corresponding Schr\"{o}dinger equation. In particular, for the large cubicities, the temperature $T$ becomes $\Lambda$ independent linearly decreasing function of the growing $\beta$ since in this limit the boundary conditions can not alter very strong $V_{eff}$. It is shown that the temperature increase, which is produced in the linear GL regime by the negative de Gennes distance, is wiped out by the growing cubicity. In this case, the decreasing $T$ passes through its bulk value $T_c$ at the unique density $n_s^{(0)}$ only, and the corresponding extrapolation length $\Lambda_{T=T_c}$ is an analytical function of $\beta$ whose properties are discussed in detail. For the densities smaller than $n_s^{(0)}$, the temperature stays above $T_c$ saturating for the large cubicities to the value determined by $n_s$ and negative $\Lambda$ while for $n_s>n_s^{(0)}$ the superconductivity is destroyed by the growing GL nonlinearity at some temperature $T>T_c$, which depends on $\Lambda$, $n_s$ and $\beta$. It is proved that the concentration $n_s^{(0)}$ transforms for the large cubicities into the density of the bulk sample.
\end{abstract}

\maketitle

\newpage
\section{Introduction}\label{sec_Intro}
For more than sixty years, the predictions based on the analysis of the Ginzburg-Landau (GL) equations \cite{Ginzburg1} for the order parameter $\Psi(\bf r)$, supercurrent density ${\bf j}(\bf r)$ and magnetic field $\bf B$
\numparts\label{GL1}
\begin{eqnarray}\label{GL1_1}
\frac{1}{2m}\left(-i\hbar{\bm\nabla}-q{\bf A}\right)^2\Psi+\beta\left|\Psi\right|^2\Psi+\alpha\Psi=0\\
\label{GL1_2}
{\bf j}=i\frac{q\hbar}{2m}\left(\Psi^*{\bm\nabla}\Psi-\Psi{\bm\nabla}\Psi^*\right)-\frac{q^2}{m}{\bf A}\left|\Psi\right|^2\\
\label{GL1_3}
{\bm\nabla}\times{\bf B}=\mu{\bf j},
\end{eqnarray}
\endnumparts
have been in a very good agreement with the experiment \cite{Ginzburg2,Ginzburg3,Moshchalkov2}. In the above equations, $m$ is a mass of a superconducting carrier, $q$ is its negative charge, $\mu$ is a magnetic permeability, $\bf A$ is a vector potential through which the magnetic field is expressed according to $${\bf B}={\bm\nabla}\times{\bf A},$$ and $\alpha$ and $\beta>0$ are GL parameters. Developed several years later microscopic Bardeen-Cooper-Schrieffer (BCS) theory \cite{Bardeen1,deGennes1,Schmidt1} had shown that the superconducting quantum (commonly called after this a Cooper pair) is actually two electrons bounded to each other by their interaction via phonons with its total mass being equal to the double electron mass $m_e$, $m=2m_e$, and negative charge of the double elementary charge $e$, $q=-2|e|$. In fact, the choice of mass in the above equations is arbitrary \cite{deGennes1} while the double negative charge reflects a fundamental fact of pairing of the two electrons with the opposite spins \cite{deGennes1}. It is postulated that the density $n_s(\bf r)$ of the superconducting electrons is determined by the order parameter $\Psi(\bf r)$ according to
\begin{equation}\label{density1}
n_s\left(\bf r\right)=\left|\Psi\left(\bf r\right)\right|^2/2.
\end{equation}
The expressions for the GL coefficients can be derived from the microscopic theory of superconductivity \cite{Gorkov1,Schmidt1}:
\begin{eqnarray}\label{CoeffAlpha1}
\alpha&=&-\frac{\hbar^2}{2m\xi^2\left(T\right)}\\
\label{CoeffBeta1}
\beta&=&\frac{1}{N\left(0\right)}\left[\frac{\hbar^2}{2m\xi^2\left(0\right)}\right]^2\frac{1}{\left(k_BT_c\right)^2}.
\end{eqnarray}
Here, $N(0)$ is the density of states at the Fermi energy, $k_B$ is Planck constant, $T_c$ is a bulk critical temperature of superconductor, and $\xi(T)$ is a coherence length varying with the temperature $T$ as
\begin{equation}\label{Xi1}
\xi\left(T\right)=\xi\left(0\right)\sqrt{\frac{T_c}{T_c-T}}.
\end{equation}
It means that the cubicity $\beta$ does not depend on the temperature while the coefficient $\alpha$ linearly grows with it, $\alpha\sim(T-T_c)$. For the 'dirty' metals (alloys), the value of $\xi^2(0)$ in \eref{CoeffAlpha1} and \eref{CoeffBeta1} has to be replaced by $\xi(0)l$ with $l$ being the mean free path of the material \cite{deGennes1,Schmidt1,Gorkov2}.

For the spatially confined samples, equations~\eref{GL1} have to be supplemented by the boundary conditions at the superconductor three dimensional (3D) surface, or 2D contour, or 1D interval $\cal S$. For the order parameter, de Gennes proposed to use the following requirement \cite{deGennes1}:
\begin{equation}\label{BoundaryCondition1}
\left.{\bf n}\left({\bm\nabla}-i\frac{q}{\hbar}{\bf A}\right)\Psi\right|_{\cal S}=\left.\frac{1}{\Lambda}\Psi\right|_{\cal S}
\end{equation}
with $\bf n$ being an inward unit vector normal to the interface. Analysis of \eref{GL1_2} and \eref{BoundaryCondition1} manifests that the real value of the extrapolation length (or the de Gennes distance, as it is customarily called on the superconductor slang) $\Lambda$ guarantees that no supercurrent flows through the confining surface\footnote{Resonant features of the supercurrent $\bf j$ crossing the interface  $\cal S$ for the nonzero imaginary part of $\Lambda$ have been predicted recently \cite{Olendski1,Olendski2} in the framework of linearized GL equations.}:
\begin{equation}\label{condition1}
\left.{\bf n}{\bf j}\right|_{{\rm Im}(\Lambda)=0}\equiv 0.
\end{equation}
Numerical magnitude of $\Lambda$ characterizes the intensity of the interaction of the superconductor with the ambient environment and, as a consequence of this, the penetration of the order parameter into it; in particular, for the border with the vacuum or the insulator its approaches infinity, takes finite positive values for the contact with the normal nonsuperconducting metal and turns to zero while bordering the magnetic material. Utilizing BCS equations, it can be shown that, for example, for the superconductor half-space terminated at $x=0$, the fading propagation of the order parameter into the neighbouring medium at $x<0$ is described as \cite{Zaitsev2,Zaitsev1}
\begin{equation}\label{Penetration1}
\Psi\left(x\right)\sim\frac{1}{\Lambda}\,e^{-\left|x\right|/\Lambda},\quad x\ll0.
\end{equation}
Note that, according to \eref{Penetration1}, the Cooper pairs can penetrate  neither into the vacuum, $\Lambda=\infty$, (what is natural) nor into the magnetics, $\Lambda=0$, as expected, since the spins of all charged carriers in the magnetic material are forced to align in the same direction while in the Cooper pair the spins of each constituent electron are antiparallel. Interestingly, for the border with the other superconductor with higher $T_c$ the extrapolation length takes negative values what physically means a surface enhancement of the superconductivity. This boundary-induced increase of the critical temperature was indeed observed in cold worked In$_{0.993}$Bi$_{0.007}$ foils \cite{Fink1} and tin samples \cite{Kozhevnikov1,Kozhevnikov2}. For the former study it was estimated from the increase of the critical temperature that the negative extrapolation length is in the order of micron \cite{Montevecchi1}, $\Lambda\approx -1$ $\mu$m, while for the latter case it was argued that the magnitude of the de Gennes distance can be controlled by, for example, manipulating the abrasive grain size and the annealing time and temperature \cite{Kozhevnikov1}. Another method of varying the extrapolation length was proposed recently from the calculation of the influence of the external electric field $\mbox{\boldmath${\cal E}$}$ on superconductors; namely, it was argued \cite{Lipavsky1,Morawetz1} that its application perpendicular to the interface transforms the zero-field de Gennes distance $\left.\Lambda\right|_{{\cal E}=0}$ into its total counterpart $\Lambda_{tot}$:
\begin{equation}\label{TotalExtrapolationLength1}
\frac{1}{\Lambda_{tot}}=\left.\frac{1}{\Lambda}\right|_{{\cal E}=0}+\frac{\cal E}{U_s}
\end{equation}
with the voltage $U_s$ being expressed through the parameters of the GL theory:
\begin{equation}\label{PotentialU}
\frac{1}{U_s}\cong\kappa^2\frac{\partial\ln T_c}{\partial\ln n_s}\frac{q\epsilon_s}{mc^2}.
\end{equation}
Here, dimensionless GL parameter $\kappa$ is the ratio of the zero-temperature London penetration length $\lambda(0)$ to the coherence length, $\kappa=\lambda(0)/\xi(0)$; $\epsilon_s$ is superconductor ionic background permittivity, and $c$ is the speed of light. While for the classical (low-$T_c$) materials with the small $\kappa$ the magnitude of $U_s$ reaches high values of $\sim10^6$ V, for the high-$T_c$ superconductors with the large GL parameter, $\kappa\sim100$, it drops significantly to the value of several tens of volts \cite{Lipavsky1}.

In the absence of the magnetic fields, equations~\eref{GL1_1} and \eref{BoundaryCondition1} simplify to
\begin{eqnarray}
\label{GL2_1}
-\frac{\hbar^2}{2m}\Delta\Psi+\beta\left|\Psi\right|^2\Psi+\alpha\Psi=0\\
\label{BoundaryCondition2}
\left.{\bf n}{\bm\nabla}\Psi\right|_{\cal S}=\left.\frac{1}{\Lambda}\Psi\right|_{\cal S}.
\end{eqnarray}
Remarkably, equation \eref{GL2_1} is formally identical to the Schr\"{o}dinger equation describing the motion of the quantum particle with the  energy
\begin{equation}\label{Energy1}
E\equiv-\alpha=\frac{\hbar^2}{2m\xi^2\left(0\right)}\left(1-\frac{T}{T_c}\right)
\end{equation}
in the domain confined by the surface $\cal S$ with the effective potential $V_{eff}(\bf r)$ of the form
\begin{equation}\label{EffectivePotential1}
V_{eff}({\bf r})=\beta\left|\Psi({\bf r})\right|^2.
\end{equation}
From \eref{Energy1} one sees that getting the lowest eigenvalue $E_{min}$ of the energy is equivalent to finding the temperature $T$ of the sample:
\begin{equation}\label{Temperature1}
T=T_c\left[1-E_{min}\frac{2m\xi^2\left(0\right)}{\hbar^2}\right].
\end{equation}
Thus, a solution of the problem of the maximal possible minimization of $E_{min}$ shows the way of increasing $T$. The procedure of determining the temperature through the lowest eigenenergy $E_{min}$ of \eref{GL1_1} and/or \eref{GL2_1} is widely used in the superconductor research \cite{deGennes1,Moshchalkov1} and will be implemented below too. Note that the effective potential $V_{eff}(\bf r)$ is always nonnegative and, accordingly, taking into account of the cubic term leads to the decrease of the temperature $T$ from \eref{Temperature1}. This has a natural physical explanation: the superconducting carriers density $n_s$, which is proportional to the square order parameter too, grows with the decrease of the temperature and, accordingly, the role of the nonlinear term becomes more prominent as well.

Armed with the experimental and theoretical knowledge discovered during last fifteen-twenty years, we readdress here the problem of the GL description of the properties of the straight superconducting film in the absence of the magnetic fields, ${\bf B}={\bf 0}$, partially discussed in the mid 1960s \cite{Zaitsev1} and in 1993 \cite{Andryushin1,Andryushin2}. Even though some basic equations for this geometry have been derived before, their analysis  in the just cited references was limited to the linear case and nonnegative $\Lambda$ only. In particular, it was shown that the critical film width $d_c$ at which the superconductivity emerges, depends on $\Lambda$ as \cite{Zaitsev1,Andryushin1,Lykov1}
\begin{equation}\label{CriticalWidth1}
d_c=2\xi\left(T\right){\rm \arctan}\frac{\xi\left(T\right)}{\Lambda}.
\end{equation}
Linearized GL theory at the negative $\Lambda$ was also discussed recently for the same configuration \cite{Montevecchi1,Montevecchi2,Slachmuylders1,Olendski3}. Below, we provide a complete analysis of the temperature $T$ dependence on the cubicity $\beta$, superdensity $n_s$ and the extrapolation length $\Lambda$. For achieving this, we analytically solve \eref{GL2_1} with the GL coefficient $\alpha$ substituted by the energy $E$, equation \eref{Energy1}, and, after imposing on its eigenfunction boundary condition \eref{BoundaryCondition2}, arrive at the transcendental equation for the calculation of the eigenvalue $E_{min}$, which, according to \eref{Temperature1}, defines the film temperature $T$. Our primary interest is an investigation of the combined influence of the extrapolation length $\Lambda$ and cubicity $\beta$ on the temperature $T$. It appears that the crucial role in this case is played also by the density $n_s$. Special attention is paid to the case of the negative de Gennes distance when, as it is proved below, the increasing cubicity promotes the energy from  the negative area to the positive values for one specific density $n_s$ only. Critical negative extrapolation length $\Lambda_0$ at which the energy turns to zero, is calculated analytically as a function of $\beta$ and its mathematical and physical analysis is performed. 

The results are presented in the following outline. In \sref{sec_Model}, a necessary formalism applied to the study of the straight GL strip is derived along with several of its asymptotic cases some of which have been known before. Results of the calculations for the positive and negative extrapolation lengths are presented separately in \sref{sec_Results}. The discussion is  wrapped up in \sref{sec_Conclusions} by some concluding remarks.

\section{Model and Formulation}\label{sec_Model}
Consider a straight 3D infinite superconducting film of width $d$. Origin of the Cartesian system of coordinates coincides with the middle of the film and the horizontal $x$ axis is perpendicular to its edges. Left (right) confining surface is characterized by the uniform along the interface de Gennes length $\Lambda_-$ ($\Lambda_+$). Frequently used method in the study of the superconducting systems implements the scaling where the order parameter $\Psi$ is expressed in units of $\sqrt{\beta/|\alpha|}$ and all distances are measured in units of the coherence length $\xi(T)=\sqrt{\hbar^2/(2m|\alpha|)}$ \cite{deGennes1,Andryushin1,Schmidt1}. Advantage of such an approach lies in the fact that no any coefficients appear in the resulting differential GL equation and the eigenvalue transcendental relationship does contain the film width $d$ \cite{Andryushin1}. However, shortcomings of this method are a continuation of its strong sides; namely, by eliminating the physical parameters from the GL equation, such scaling shadows the explicit interrelations between them  and at the final stage one needs to return again to the units where these quantities re-emerge what, due to the quite complicated structure of the derived dependencies, is not an easy task to do \cite{Andryushin1}. For example, in this approach the parameter $\beta$ is dropped out from the integration constants of the GL equation and, accordingly, it is not clear how to explicitly obtain the influence of its magnitude on the properties of the superconductor and to track the transformation from the linear to the cubic scenario. Since in the present research we are interested in the temperature dependence on, among other parameters, the cubicity $\beta$, we employ the scaling where the distances are measured in units of the film width $d$, which is the only natural (temperature independent) length for the geometry under consideration. Accordingly, the energy $E$ through which the temperature $T$ is expressed according to \eref{Temperature1}, will be measured in units of the ground state energy $\pi^2\hbar^2/(2md^2)$ of the 1D Dirichlet well of the width $d$, and cubicity $\beta$ will be expressed in units of $d^2$. In this way, the temperature $T$ (via the energy $E$) and cubicity $\beta$ explicitly stay in the differential equation and enter its solution and the resulting transcendental equation what allows to directly investigate the correlations between these two parameters.

In the absence of the magnetic field, ${\bf B}={\bf 0}$, the motion along each of the axes is independent from the movement along the other ones. Mathematically, this separation of the transverse and two longitudinal motions in the film is expressed via the following ansatz:
\begin{equation}\label{ansatz1}
\Psi\left(x,y,z\right)=e^{i\left(k_yy+k_zz\right)}\chi\left(x\right).
\end{equation}
After substitution of \eref{ansatz1} into \eref{GL2_1} one arrives at the differential equation for the transverse function $\chi(x)$:
\begin{equation}\label{ChiEq2}
\chi''\left(x\right)+\left(\pi^2E-k_y^2-k_z^2\right)\chi\left(x\right)-\beta\chi^3\left(x\right)=0.
\end{equation}
The designations of the absolute values in the cubic term have been dropped out from \eref{ChiEq2} since it is known that even in the presence of the magnetic fields a solution of the GL equations in the simply connected region can be chosen real \cite{Schmidt1}.  As we are interested in the lowest possible value of the energy that, according to \eref{Temperature1}, guarantees the highest attainable temperature $T$, we put $k_y$ and $k_z$ equal to zero to obtain:
\begin{equation}\label{ChiEq1}
\chi''\left(x\right)+\pi^2E\chi\left(x\right)-\beta\chi^3\left(x\right)=0.
\end{equation}
Second-order nonlinear differential equation  \eref{ChiEq1} is supplemented by the two boundary conditions at the film edges:
\begin{equation}\label{Boundary3}
\left.\left[\chi'\left(x\right)\pm\frac{1}{\Lambda_\pm}\chi\left(x\right)\right]\right|_{x=\pm 1/2}=0.
\end{equation} 
Below, for simplicity, we will use the same extrapolation lengths at the opposite surfaces, 
\begin{equation}\label{EqualLengths1}
\Lambda_+\equiv\Lambda_-\equiv\Lambda.
\end{equation}

General solution to \eref{ChiEq1} reads:
\begin{equation}\label{Solution1}
\chi\left(x\right)=\sqrt{\frac{2}{\beta}}\,\eta\zeta\,{\rm sn}\!\left(\zeta x+C,\eta\right)
\end{equation} 
with
\begin{eqnarray}
\label{Coef1_eta}
\eta&\equiv&\eta\left(E,\beta,k\right)=\sqrt{\frac{\beta}{2\pi^2E-\beta}}\,k,\\
\label{Coef1_zeta}
\zeta&\equiv&\zeta\left(E,\eta\right)=\sqrt{\frac{\pi^2 E}{1+\eta^2}}.
\end{eqnarray}
Here, ${\rm sn}(u,k)$ is a Jacobi elliptic sine \cite{Gradshteyn1,Whittaker1}, the integration factor $k$ is determined by (or, vice versa, determines) the density $n_s$ of the Cooper pairs, as it directly follows from comparison between \eref{Solution1}, \eref{Coef1_eta} and \eref{density1}, and the coefficient $C$ is found from the symmetry properties of the film; namely, for the equal de Gennes lengths, equation \eref{EqualLengths1}, one has either even, $\chi_{ev}\left(-x\right)=\chi_{ev}\left(x\right)$, or odd, $\chi_{odd}\left(-x\right)=-\chi_{odd}\left(x\right)$, solutions. Recalling properties of the elliptic functions \cite{Gradshteyn1,Whittaker1,Abramowitz1}, one immediately gets:
\numparts\label{CoeffC1}
\begin{eqnarray}
\label{CoeffCeven}
C_{ev} &=&{\bf K}(\eta)\\
\label{CoeffCodd}
C_{odd} &=& 0,
\end{eqnarray}
\endnumparts
where ${\bf K}(k)$ is a complete elliptic integral of the first kind \cite{Gradshteyn1,Whittaker1,Abramowitz1}. Accordingly, the order parameter becomes
\numparts\label{Solution2}
\begin{eqnarray}\label{EvenSolution}
\chi_{ev}\left(x\right)=\sqrt{\frac{2}{\beta}}\,\eta\zeta\,{\rm cd}\!\left(\zeta x,\eta\right)\\
\label{OddSolution}
\chi_{odd}\left(x\right)=\sqrt{\frac{2}{\beta}}\,\eta\zeta\,{\rm sn}\!\left(\zeta x,\eta\right)
\end{eqnarray}
\endnumparts
with ${\rm cd}(u,k)$ being another Jacobi elliptic function (we follow commonly adopted \cite{Gradshteyn1,Whittaker1,Abramowitz1,Bateman1} Glaisher designations \cite{Glaisher1,Glaisher2}). Applying boundary conditions \eref{Boundary3} to the transverse function $\chi(x)$, one arrives at
\numparts\label{EigenvalueEq1}
\begin{eqnarray}\label{EigenvalueEq1_Even}
\zeta\left(1-\eta^2\right){\rm sn}\!\left(\zeta/2,\eta\right)-\frac{1}{\Lambda}\,{\rm cn}\!\left(\zeta/2,\eta\right){{\rm dn}\!\left(\zeta/2,\eta\right)}=0\\
\fl{\rm for\,\,the\,\,even\,\,states,\,\,and}\nonumber\\
\label{EigenvalueEq1_Odd}
\zeta{\rm cn}\!\left(\zeta/2,\eta\right){\rm dn}\!\left(\zeta/2,\eta\right)+\frac{1}{\Lambda}\,{\rm sn}\!\left(\zeta/2,\eta\right)=0
\end{eqnarray}
\endnumparts
for the odd ones with ${\rm cn}(u,k)$ and ${\rm dn}(u,k)$ being another two major elliptic functions \cite{Gradshteyn1,Whittaker1,Abramowitz1}. These equations determine energy $E$ as a function of the de Gennes distance $\Lambda$ and, through the factor $\eta$, of the cubicity $\beta$ and parameter $k$: $E\equiv E\left(\Lambda,\beta,k\right)$. It is the aim of the next chapter to analyze these dependencies for the positive, zero and negative extrapolation lengths. Since the lowest energy level determining, according to \eref{Temperature1}, the temperature $T$ of the sample, is always even, below mainly this case is discussed. 
It is important to state that the mathematics involved in the derivation of (26) and (27) automatically
introduces another important physical parameter; namely, it is the integration constant $k$ that plays, as we shall see below, a crucial role in determining temperature of the film  and a square of which, as pointed out above, represents the density $n_s$. Note that the coefficients $\beta$ and $k$ are not completely independent, as it directly follows from \eref{Coef1_eta}, 
\eref{Coef1_zeta} and (27); for example, if either of them is zero, equations (27) do not depend on the
second one. Mathematically, this is explained by the dependence of the effective potential $V_{eff}(x)$ from \eref{EffectivePotential1} on $k$ via the order parameter $\chi(x)$. This has a clear physical meaning too; namely, for the linear case, $\beta=0$, the density of the superconducting pairs is vanishingly small \cite{deGennes1} what corresponds to the coefficient 
$k$ tending to zero, $k\rightarrow 0$, and, as a result, it is dropped out from (27). Below we will
interchangeably operate with the factor $\eta$ or the parameters $\beta$ and $k$ keeping in mind that, if required, the coefficient $k$, for example, for each given $\beta$ can be determined from \eref{Coef1_eta} after finding a solution $E$ of equations 
(27). Utilizing properties of the Jacobi elliptic functions \cite{Gradshteyn1,Abramowitz1,Bateman1}, it can be shown
that the transformation $\eta\rightarrow1/\eta$ leaves \eref{EigenvalueEq1_Odd} intact while its even counterpart \eref{EigenvalueEq1_Even} changes sign of $\Lambda$ what means that it is sufficient to consider the unit interval
\begin{equation}\label{Limit1}
0\le\eta\le 1.
\end{equation}
At its right edge, equation~\eref{EigenvalueEq1_Even} becomes:
\begin{equation}\label{EigenvalueEq1_EvenRightLimit}
ue^u=\frac{4}{\Lambda}\frac{1}{1-\eta^2}
\end{equation}
wit $u=\pi\sqrt{E/2}$ and, thus, the energy $E$ in this limit is expressed  via the Lambert $W$ function \cite{Corless1}
\begin{equation}\label{Lambert1}
E=\frac{2}{\pi^2}W^2\left(\frac{4}{\Lambda}\frac{1}{1-\eta^2}\right),\quad\eta\rightarrow1.
\end{equation}
Alternatively, one can use an  iterative solution of \eref{EigenvalueEq1_EvenRightLimit} diverging to infinity as
\begin{equation}\label{Limit2}
\fl E=\frac{2}{\pi^2}\ln^2\!\left(-\frac{\Lambda}{4}\left(1-\eta^2\right)\ln\!\left(-\frac{\Lambda}{4}\left(1-\eta^2\right)\ln\!\left(\frac{\Lambda}{4}\left(1-\eta^2\right)\right)\right)\right),\quad\eta\rightarrow 1,
\end{equation}
where the number of the nested terms $-\frac{\Lambda}{4}\left(1-\eta^2\right)\ln\left(-\frac{\Lambda}{4}\left(1-\eta^2\right)\ln(\ldots)\right)$ can be increased or decreased due to the desired precision. Expansion \eref{Limit2} also follows straightforwardly from the asymptotic properties of the Lambert $W$ function \cite{Corless1}. Of course, in reality the energy grows until the temperature $T$ from \eref{Temperature1} turns to zero. In terms of $\beta$ and $k$, the leading, $\Lambda$-independent coefficient of \eref{Limit2} can be deduced directly from \eref{Coef1_eta}:
\begin{equation}\label{Limit6}
E\rightarrow\frac{1+k^2}{2\pi^2}\beta,\quad k\,\,{\rm or/and}\,\,\beta\rightarrow\infty.
\end{equation}
The corresponding order parameter tends then to the $\beta$-independent constant:
\begin{equation}\label{Chi2}
\chi(x)=\sqrt{\frac{1+k^2}{2}},\quad k\,\,{\rm or/and}\,\,\beta\rightarrow\infty.
\end{equation}
Note that at $k=1$ and very large nonlinearities, \eref{Limit6} and \eref{Chi2} constitute a solution of the  initial normalized GL equation \eref{ChiEq1} for the bulk superconductor when no any boundaries are present:
\numparts\label{Special1}
\begin{eqnarray}\label{Limit5}
\left.E\right|_{k=1}\rightarrow\frac{\beta}{\pi^2},\quad\beta\rightarrow\infty,\\
\label{Chi3}
\left.\chi(x)\right|_{k=1}=1,\quad\beta\rightarrow\infty.
\end{eqnarray}
\endnumparts
This means that the effective potential \eref{EffectivePotential1} is so large in this limit that the confinement of the strip can not alter it. Equations~(34)  also show that the density with $k=1$ is a special one what will be vividly demonstrated for the negative de Gennes distances, sub\sref{sec_ResNegative}.

In the opposite limit of the zero $\eta$ one arrives at the equations defining energies $E$ in the linear regime \cite{Slachmuylders1,Olendski3,AlHashimi1}, as expected. Other analytical asymptotics of the above equations is the case of the Neumann boundary condition, $1/\Lambda=0$, when the ground-state energy is an identical zero for any $\beta$ and $k$, as it follows directly from \eref{EigenvalueEq1_Even}: 
\begin{equation}\label{Limit3}
E(\Lambda=\infty,\beta,k)\equiv 0.
\end{equation}
It means that for the Neumann strip the critical temperature does not depend on the cubicity of the GL equation and on the density of the superconducting carriers and is always equal to the bulk critical temperature $T_c$. As a final remark of this section, let us note that if one chooses as a unit of length not the strip width $d$ but the coherence length $\xi(T)=\sqrt{\hbar^2/(2mE)}$, then the changes  $\sqrt{\pi^2E}\rightarrow d$ in  the first argument of the elliptic functions and $\sqrt{\pi^2E}\rightarrow 1$  in all other 
appearances in (27) of the parameter $\zeta$ from \eref{Coef1_zeta} allows one to find the critical width
$d_c$ as a function of the de Gennes distance $\Lambda$ and factor $\eta$ \cite{Andryushin1}. However, compared to the normalization implemented by us above, the cubicity $\beta$ in such scaling does disappear from the corresponding equations since the role of the primary parameter in this case is played not by $k$ but directly by $\eta$ that emerges as a result of the integration of the coefficient-free GL equation \cite{Andryushin1}
\begin{equation}\label{ChiEq4}
\tilde{\chi}''\left(x\right)+\tilde{\chi}\left(x\right)-\tilde{\chi}^3\left(x\right)=0,
\end{equation}
where the tilde denotes that the order parameter is measured in units of $\sqrt{\beta/|\alpha|}$ \cite{Andryushin1}.

\section{Results and discussion}\label{sec_Results}
Here, the results of the calculations based on the theory developed in the previous section are presented and analysed from the mathematical and physical points of view. Since the obtained dependencies are different for the opposite signs of the de Gennes distance, the cases of $\Lambda\ge0$ and $\Lambda<0$ are considered separately.
\begin{figure}
\centering
\includegraphics[width=0.99\columnwidth]{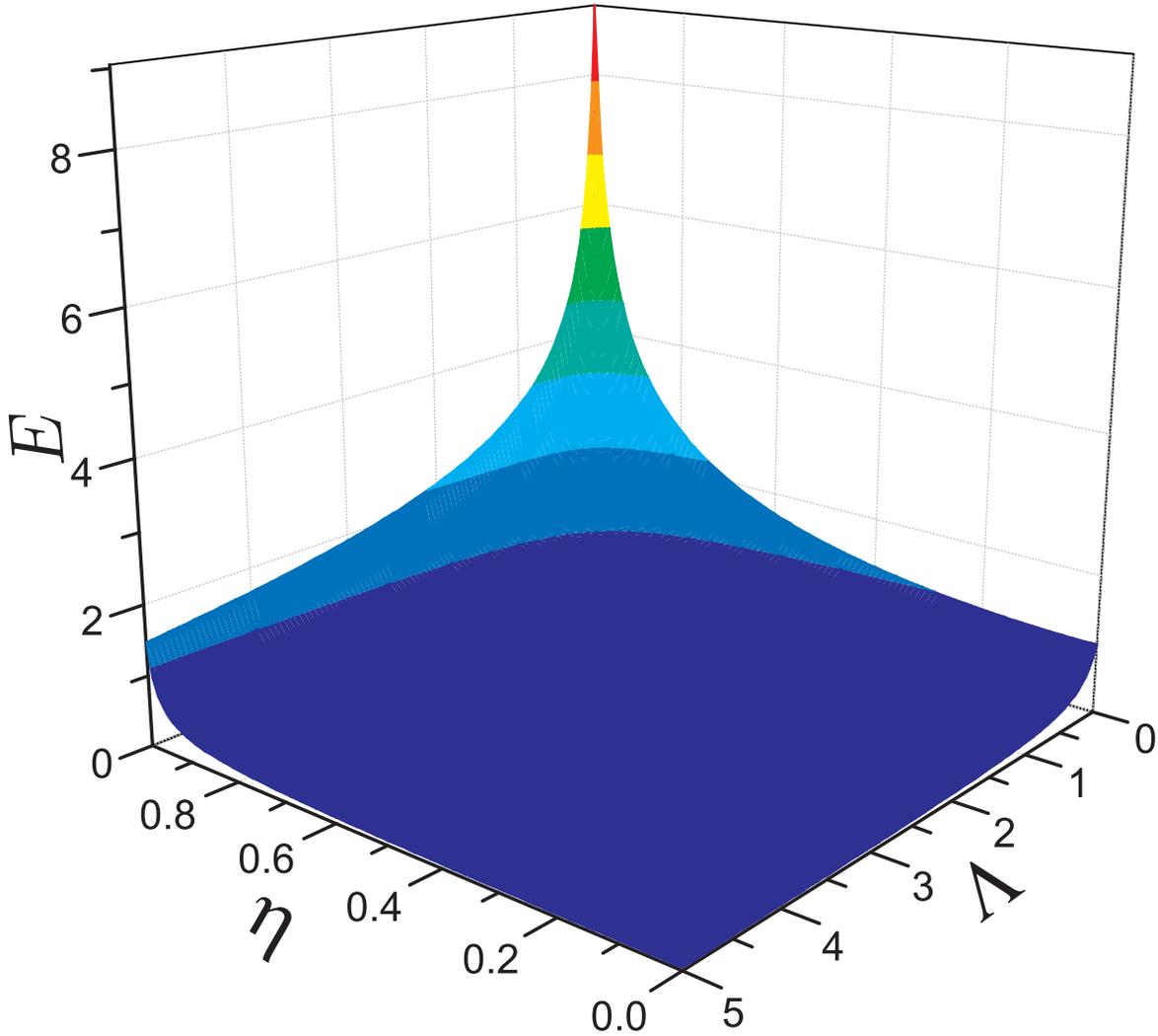}
\caption{\label{Fig1}
Lowest energy $E$ as a function of the coefficient $\eta$ and positive de Gennes distance $\Lambda$. At $\eta$ approaching unity, the energy tends to infinity according to \eref{Limit2}.
}
\end{figure}

\subsection{Zero and positive extrapolation lengths}\label{sec_ResPositive}
\Fref{Fig1} shows energy $E$ as a function of the parameter $\eta$ and positive de Gennes distance $\Lambda$. Similar to the linear configuration \cite{Olendski3}, for all nonzero $\eta$ the energy decreases with the growth of the extrapolation length until in the Neumann limit, $1/\Lambda=0$, one gets the critical temperature of the film equal to its bulk counterpart, as it was stated in the previous section, equation~\eref{Limit3}. In turn, the growth of $\eta$ pushes the energy upwards and at $\eta\rightarrow1$ one sees the divergence described by \eref{Lambert1} and/or \eref{Limit2} with its steepness being larger for the smaller $\Lambda$.
\begin{figure}
\centering
\includegraphics[width=0.6\columnwidth]{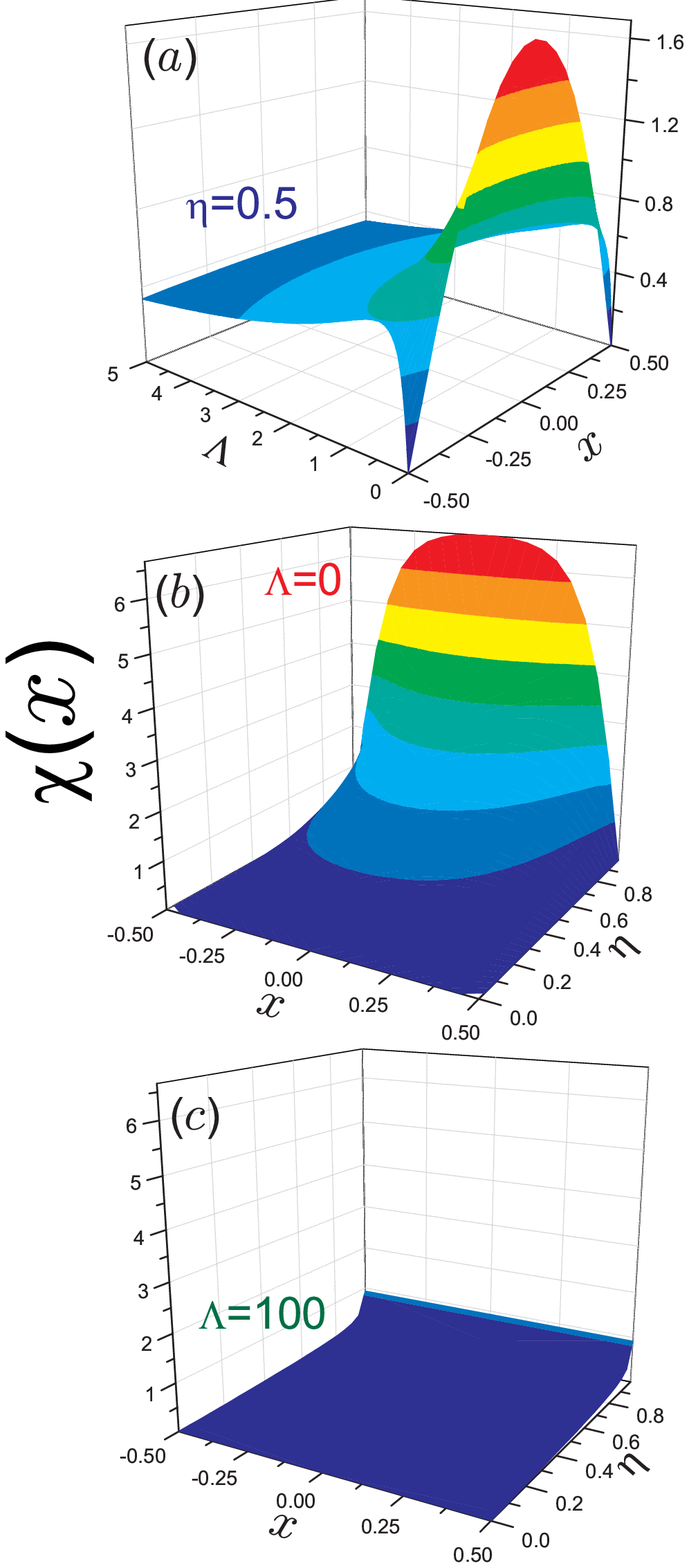}
\caption{\label{Fig2}
Order parameter $\chi(x)$ [normalized to $(2/\beta)^{1/2}$] as a function of (a) the de Gennes length $\Lambda$ for $\eta=0.5$, and of the parameter $\eta$ for (b) $\Lambda=0$ and (c) $\Lambda=100$. Note different vertical scale in panel (a) as compared to panels (b) and (c).
}
\end{figure}

Evolution of the order parameter is shown in \fref{Fig2} where the panel (a) exhibits  $\chi(x)$ with varying the de Gennes distance at the fixed $\eta=0.5$. Similar dependencies are characteristic for the other factors $\eta$ too. For the Dirichlet case, $\Lambda=0$, due to the suppression of the  order parameter at the edges, the nucleation of the superconductivity takes place in the middle of the strip. Growing extrapolation length allows the Cooper pairs to approach the interfaces with the corresponding decrease of the slope of the order parameter in the transverse direction, and, at the large $\Lambda$, the function $\chi(x)$ almost does not depend on $x$ what means a uniform nucleation of the superconductivity across the film.

Panels (b) and (c) of \fref{Fig2} provide a comparative analysis of the influence of the factor $\eta$ on the order parameter at the two extreme values of the extrapolation length: small, $\Lambda=0$, panel (b), and very large, $\Lambda=100$, panel (c). In either case, the symmetric function $\chi(x)$ unrestrictedly grows when $\eta$ approaches unity. However, the $\chi$-shape for the  Dirichlet case is much more convex as compared to the de Gennes distances close to the Neumann one, $1/\Lambda\sim 0$, when, as panel (c) demonstrates, the order parameter stays flat with its magnitude close to zero for almost all values of $\eta$ and starts to grow upwards only in the very neighbourhood of $\eta=1$. The larger the de Gennes distance is, the sharper the rise is at the right edge of the interval from \eref{Limit1}.

\begin{figure}
\centering
\includegraphics[width=0.99\columnwidth]{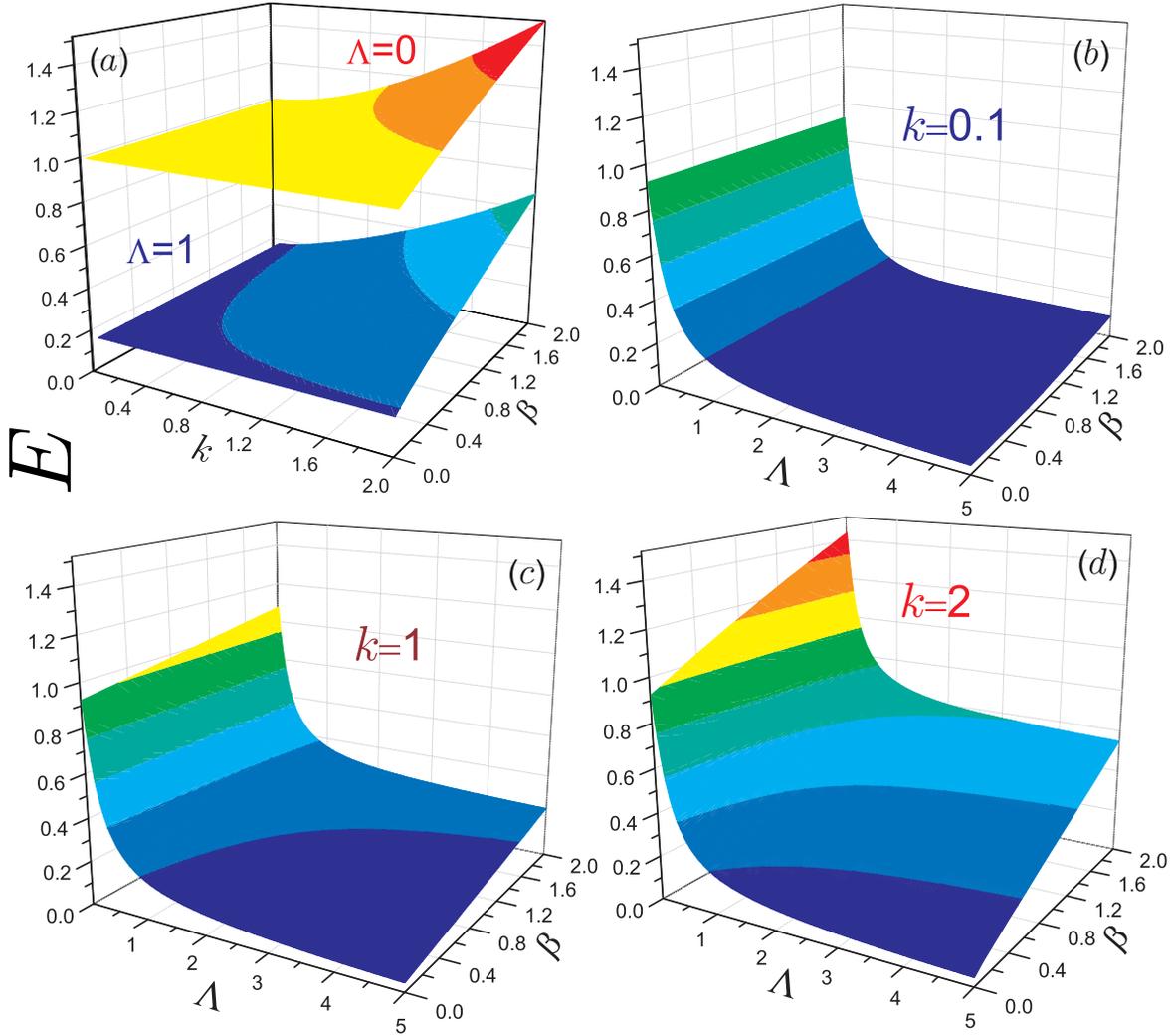}
\caption{\label{Fig3}
a) Lowest energy $E$ as a function of cubicity $\beta$ and coefficient $k$ for the Dirichlet case (upper surface) and for $\Lambda=1$ (lower surface). (b)-(d) Energy $E$ as a function of the cubicity $\beta$ and positive extrapolation length $\Lambda$ for several values of the parameter $k$ shown in each of the panels.
}
\end{figure}

According to \eref{Coef1_eta}, the index $\eta$ accommodates two variables $\beta$ and $k$.  To illustrate their separate influence on the temperature $T$, we plot in panel (a) of \fref{Fig3} the energy $E$ in terms of these two factors for the extrapolation lengths $\Lambda=0$ and $\Lambda=1$. It is seen that, as discussed in the previous section, for the zero value of either $\beta$ or $k$ the temperature does not depend on the second parameter. The energy increases and, accordingly, the temperature $T$ decreases with the growth of $\beta$ or/and $k$ for any finite de Gennes distance. As stated in the Introduction, the temperature dependence on the cubicity is explained by the positiveness of the effective potential $V_{eff}(x)$, equation \eref{EffectivePotential1}. In other words, the role of the nonlinearity increases with the temperature decreasing. The energy growth with the factor $k$ is due to the fact that the decreasing temperature $T$ leads to the increase of the density of the Cooper pairs that, as it follows from \eref{density1}, \eref{Solution1} and \eref{Coef1_eta}, is determined, among other parameters, by $k$. As the calculations show, the rate of the energy growth along the positive $k$ direction exceeds the one along the $\beta$ axis:
\begin{equation}\label{Condition5}
\frac{\partial E}{\partial k}>\frac{\partial E}{\partial\beta}.
\end{equation}
These two speeds of the energy change grow bigger with $k$ and $\beta$.

As stated in the previous paragraph, the cubicity plays more prominent role when the temperature decreases. A natural question arises: when does this role become so dominant that the linear theory could not be applied and its predictions are not close anymore to the results obtained from the cubic GL equation? Practical importance of this problem lies in the fact that obtaining the solution of the linear equation is, generally, much easier than its cubic counterpart. To answer this question, in panels (b)-(d) of \fref{Fig3} the energy is plotted in terms of the extrapolation length $\Lambda$ and cubicity $\beta$ at several parameters $k$. It is seen that for the small densities, say, $k=0.1$ in panel (b), the energy only weakly depends on the cubicity and, accordingly, the linear theory serves as a reliable substitution of the  cubic GL equation in a wide range of the parameter $\beta$.   This is  especially true for the Dirichlet case, since the energy change with the cubicity decreases for the smaller de Gennes distances:
\begin{equation}\label{Condition6}
\frac{\partial}{\partial\beta}E\!\left(\Lambda_1,\beta,k\right)>\frac{\partial}{\partial\beta}E\!\left(\Lambda_2,\beta,k\right),\quad\Lambda_1>\Lambda_2\ge0.
\end{equation}
Growing density $n_s$ leads to the increase of the slope of the $E$-$\beta$ characteristics, as a transition from panel (b) to (c) and, subsequently, to (d) demonstrates. \Eref{Condition6} holds true but the energy difference for the different extrapolation lengths gets smaller:
\begin{equation}\label{Condition7}
\frac{\partial^2}{\partial\beta\partial\Lambda}E\!\left(\Lambda,\beta,k_1\right)<\frac{\partial^2}{\partial\beta\partial\Lambda}E\!\left(\Lambda,\beta,k_2\right),\quad k_1>k_2.
\end{equation}
As a result, the interval of change of the coefficient $\beta$ where the solution of the cubic GL equation can be replaced by its linear counterpart without noticeable loss of precision, shrinks and becomes less dependent on the de Gennes distance with growing density of the superconducting carriers $n_s$. In fact, this conclusion is not unexpected since, as it was already stated above, the density $n_s$ is proportional to the square of the order parameter, equation \eref{density1}, that, in turn, is defined by the factor $k$. At the same time, the same expression $\chi^2(x)$ enters the effective potential $V_{eff}(x)$, equation \eref{EffectivePotential1}, what means that an increase of the superconducting density leads to the stronger deviation from the linear regime. Let us note also that \eref{Condition7} leads to
\begin{equation}\label{Condition8}
\frac{\partial^3}{\partial\beta\partial k\partial\Lambda}E\!\left(\Lambda,\beta,k\right)<0,
\end{equation}
what is a consequence of the negative derivative with respect to the de Gennes distance and positiveness of the partial derivatives with respect to the cubicity and density, as it is exemplified in \fref{Fig3}. Let us point out also that the approach to the limit from \eref{Limit6} for the large $\beta$ is also clearly seen from panels (b)-(d) of \fref{Fig3}.

\subsection{Negative de Gennes lengths}\label{sec_ResNegative}
Before discussing the interaction of the negative de Gennes distance and cubicity, one needs to recall the linear limit, $\beta=0$, of the same geometry \cite{Montevecchi1,Montevecchi2,Slachmuylders1,Olendski3,AlHashimi1}. In this case, the ground-state energy for $\Lambda<0$ always lies below zero. The adjacent upper state possesses negative energy too but only if the condition
\begin{equation}\label{Condition3}
\Lambda_++\Lambda_-+1<0
\end{equation}
is satisfied \cite{Olendski3}. Absolute values of the negative energies increase with dwindling $|\Lambda_\pm|$, and at the vanishing and, in general, different lengths $\Lambda_\pm$ they diverge as \cite{Olendski2}:
\begin{equation}\label{Robin1}
E_\pm=-\frac{1}{\pi^2\Lambda_\pm^2},\quad \Lambda_+\rightarrow -0,\, \Lambda_-\rightarrow -0.
\end{equation}
For the equal, equation \eref{EqualLengths1}, small negative de Gennes distances one has two almost degenerate states with their symmetric and antisymmetric order parameters localized mainly at the interfaces. For the smaller $|\Lambda|$ the attraction of the interfaces gets stronger \cite{Olendski3,AlHashimi1}, and at $\Lambda\rightarrow -0$ each of these two order parameters transforms into the corresponding superposition of the $\delta$-functions with their origins at the  interfaces \cite{AlHashimi1}:
\begin{equation}\label{OrderCritical1}
\chi_{\!\!\tiny\begin{array}{l}ev\\odd\end{array}}\!\!(x)\sim\delta(x-1/2)\pm\delta(x+1/2),\quad\Lambda\rightarrow -0.
\end{equation}
Thus, the negative extrapolation length transforms the interface into the attractive center what results in lowering the Cooper pair energy and its localization at the edge. The smaller the magnitude of the negative de Gennes distance is, the larger the attraction and the stronger the localization of the surface state are. Unlimited growth of the magnitude of the negative energy from \eref{Robin1} is extremely interesting from the practical point of view since it means, according to \eref{Temperature1}, an unrestricted growth of the temperature $T$. In the model of equations~\eref{TotalExtrapolationLength1} and \eref{PotentialU}, the limit of the infinitely small negative $\Lambda_{tot}$ can be achieved, for example, by applying to the zero-field Neumann edge an appropriately directed weak electric field $\mbox{\boldmath${\cal E}$}$. We also mention that \eref{Robin1} is a particular case of a more general property of the vanishingly small negative extrapolation length in any number of dimensions intensively studied recently by mathematicians \cite{Lacey1,Lou1,Levitin1,Daners1,Colorado1}.

As a first step in the analysis of the combined influence of the negative de Gennes distance $\Lambda$ and cubicity $\beta$ on the temperature $T$, let us note that, as it directly follows from the transformation properties of the elliptic functions with complex argument and/or parameter \cite{Gradshteyn1,Whittaker1}\footnote{There is a typo in the English translation of Ref. \cite{Gradshteyn1}; namely, the numerator of the right-hand side of equation 8.153.8 instead of the Jacobi elliptic sine should contain an elliptic cosine, as is the case in the Russian original.},
equations \eref{Solution1} and (27) remain valid for the negative energies $E$ too. Alternatively, one can directly use \eref{ChiEq1} where for $E<0$ it is necessary to put $E=-|E|$. After some algebra one gets the order parameter in the form
\begin{equation}\label{SolutionNegative1}
\chi_-\left(x\right)=\sqrt{\frac{2}{\beta}}\,\eta_-\zeta_-\,{\rm nd}\!\left(\zeta_-x,\sqrt{1+\eta_-^2}\right),
\end{equation}
where the factors $\eta_-$ and $\zeta_-$ are given as
\begin{eqnarray}
\label{Coef2_eta}
\eta_-&\equiv&\eta_-(E,\beta,k)=\sqrt{\frac{\beta}{2\pi^2\left|E\right|+\beta}}\,k,\\
\label{Coef2_zeta}
\zeta_-&\equiv&\zeta_-\left(E,\eta\right)=\sqrt{\frac{\pi^2\left|E\right|}{1-\eta_-^2}}
\end{eqnarray}
with their subscripts indicating that they correspond to the negative energies. Eigenvalue equation \eref{EigenvalueEq1_Even} turns to
\begin{equation}\label{EigenvalueNegative1}
\zeta_-\left(1+\eta_-^2\right){\rm cn}\!\left(\zeta_-/2,\eta_-\right){\rm sn}\!\left(\zeta_-/2,\eta_-\right)+\frac{1}{\Lambda}\,{\rm dn}\!\left(\zeta_-/2,\eta_-\right)=0.
\end{equation}
Note that in this case the parameter $\eta_-$ strictly varies between zero and unity:
\begin{equation}\label{Limit4}
0\le\eta_-\le1.
\end{equation}
\begin{figure}
\centering
\includegraphics[width=0.99\columnwidth]{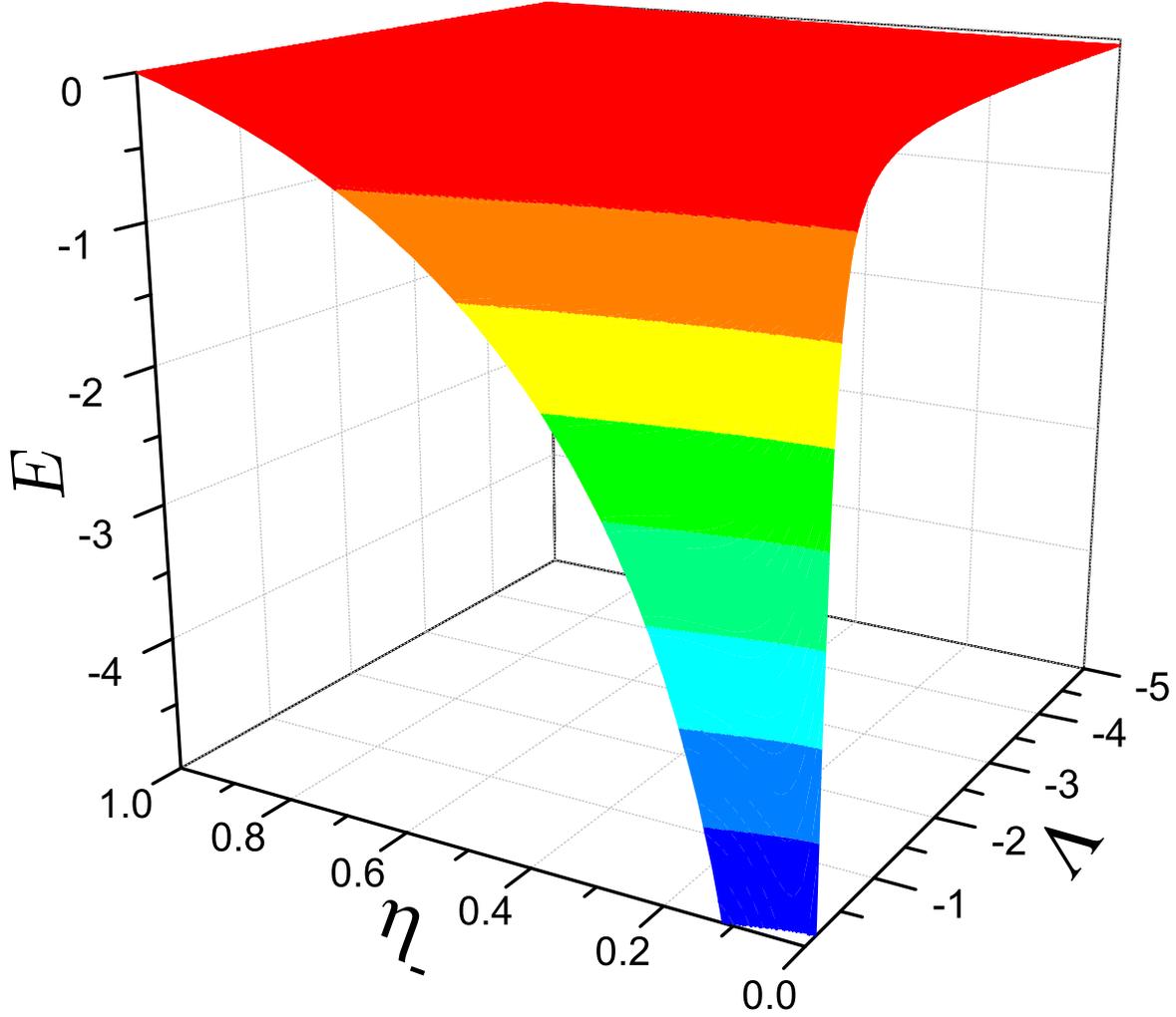}
\caption{\label{Fig4}
Energy $E$ as a function of the index $\eta_-$ and negative extrapolation length $\Lambda$.
}
\end{figure}

\begin{figure}
\centering
\includegraphics[width=0.99\columnwidth]{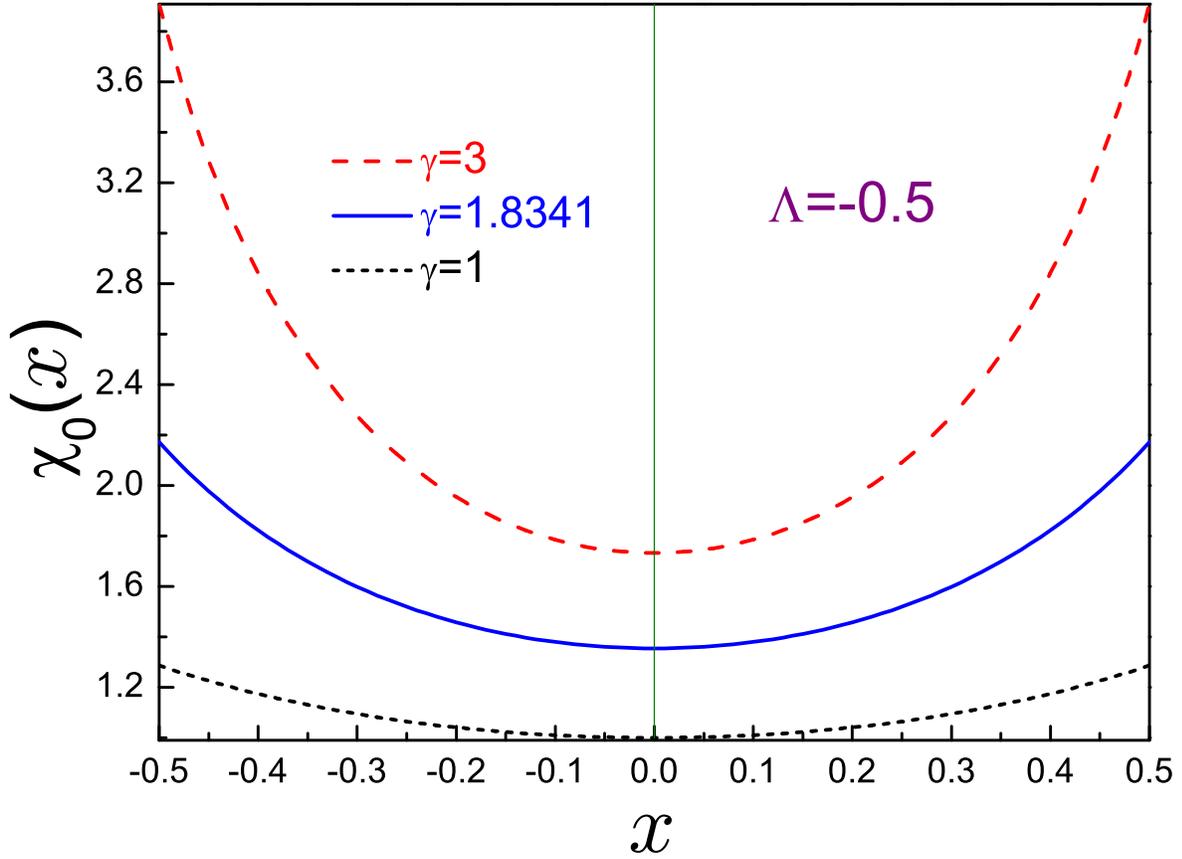}
\caption{\label{Fig5}
Order parameter $\chi_0(x)$ (in units of $\sqrt{2/\beta}$) from \eref{SolutionZero1} for $\Lambda=-0.5$ and several values of the parameter $\gamma$ where the solid line is for $\gamma=1.8341\ldots$ being a solution of \eref{Equation1}, the dotted line is for $\gamma=1$, and the dashed line is for $\gamma=3$. Thin vertical line denotes a middle of the film.
}
\end{figure}

\Fref{Fig4} depicts energy $E$ as a function of the index $\eta_-$ and negative de Gennes length $\Lambda$. In the linear regime, $\eta_-=0$, a peculiarity from \eref{Robin1} is clearly seen at the vanishing $\Lambda$. Rising factor $\eta_-$ leads to the corresponding increase of the energy and in this way withstands the temperature growth produced by the negative $\Lambda$. Its pace of change becomes faster for the smaller $|\Lambda|$:
\begin{equation}\label{Condition2}
\frac{\partial E\!\left(\Lambda_1,\eta_-\right)}{\partial\eta_-}>\frac{\partial E\!\left(\Lambda_2,\eta_-\right)}{\partial\eta_-},\quad|\Lambda_1|<|\Lambda_2|.
\end{equation}
At $\eta_-=1$ the energy turns to zero what means, according to \eref{Temperature1}, a complete compensation of the influence of the negative length, 
\begin{equation}\label{Condition4}
\left.T\right|_{\,\eta_-=1}=T_c.
\end{equation}
This equation holds for all extrapolation lengths. However, even though the de Gennes distance does not enter \eref{Condition4} explicitly, the temperature dependence on $\Lambda$ is crucial here. To prove this, we state that $\eta_-=1$ is one of the points on the semi infinite  strip of the unit width $\left\{0\le\eta_-\le1,-\infty<\Lambda<0\right\}$ in the neighbourhood of which the energy $E$ as a function of the index $\eta_-$ and the length $\Lambda$ is represented in the separable form:
\begin{equation}\label{FunctionE}
E(\eta_-,\Lambda)=\frac{1}{\pi^2}\left(\eta_-^2-1\right)\gamma\left(\Lambda\right),\quad \eta_-\rightarrow 1,
\end{equation}
where, according to \eref{EigenvalueNegative1}, the value of $\gamma$ is determined from equation
\begin{equation}\label{Equation1}
f(\gamma,\Lambda)=0
\end{equation}
with
\begin{equation}\label{FunctionF}
\fl f(y,\Lambda)=2y^{1\left/2\right.}\,{\rm cn}\!\left(\frac{1}{2}\,y^{1\left/2\right.},2^{1\left/2\right.}\!\right){\rm sn}\!\left(\frac{1}{2}\,y^{1\left/2\right.},2^{1\left/2\right.}\!\right)+\frac{1}{\Lambda}\,{\rm dn}\!\left(\frac{1}{2}\,y^{1\left/2\right.},2^{1\left/2\right.}\!\right).
\end{equation}
Then, from \eref{SolutionNegative1} and \eref{FunctionE}, the order parameter $\chi_0(x)$ for $T=T_c$ is written as
\begin{equation}\label{SolutionZero1}
\chi_0\left(x\right)=\sqrt{\frac{2}{\beta}}\,\gamma^{1\left/2\right.}{\rm nd}\!\left(\gamma^{1\left/2\right.}x,2^{1\left/2\right.}\right).
\end{equation}
Any other values of $\gamma(\Lambda)$ from \eref{FunctionE} that do not satisfy \eref{Equation1} are unphysical ones. To exemplify this, in \fref{Fig5} we plot the order parameter, equation \eref{SolutionZero1}, for several values of $\gamma$ with the solid line representing the case when \eref{Equation1} is satisfied. Similar to the linear regime discussed above, $\eta_-=0$, the probability of finding the Cooper pairs at the edges is larger than in the middle of the film. For the nonzero cubicity, $\eta_->0$, the effective potential from \eref{EffectivePotential1} pushes the energy upwards and only for the value of $\gamma$ satisfying \eref{Equation1} it exactly compensates the attractive influence of the interfaces with the negative $\Lambda$. The potential $V_{eff}(x)$ for any other $\gamma$ is either too shallow to promote the energy to zero (see the dotted line in \fref{Fig5}) or too strong (dashed line) with the corresponding temperature $T$ lying already below its bulk critical counterpart $T_c$ what in either case means that the temperature $T(\eta_-,\Lambda)$ in the vicinity of $\eta_-=1$ is {\em not} a continuous function what, obviously, eliminates these two unphysical $\gamma$ from the consideration.

\begin{figure}
\centering
\includegraphics[width=0.99\columnwidth]{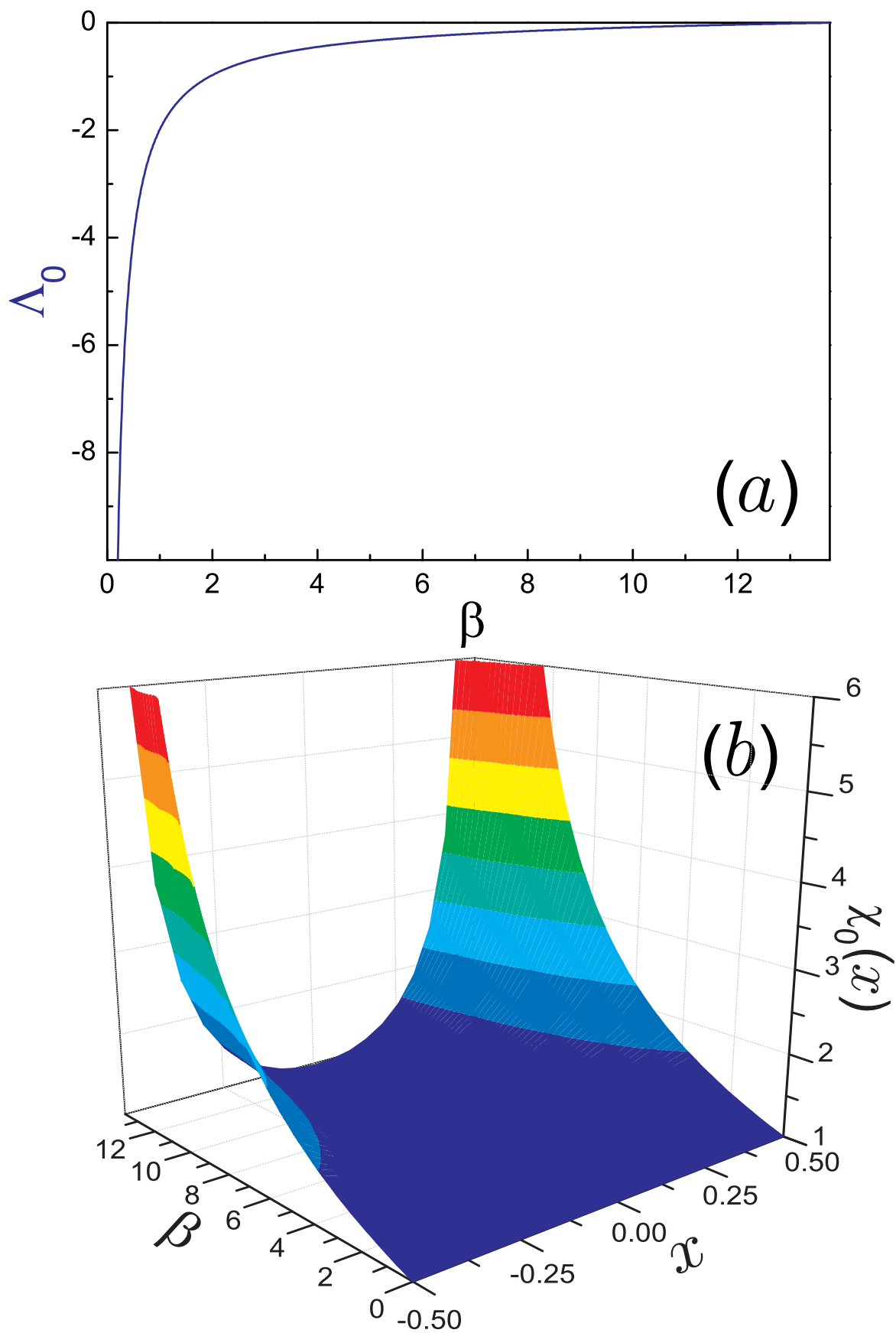}
\caption{\label{Fig6}
(a) Critical extrapolation length $\Lambda_0$, equation \eref{LambdaZeroEnergy1}, and (b) corresponding to it zero-energy order parameter from \eref{SolutionZero2} versus the cubicity $\beta$.
}
\end{figure}

Similar to the case of the positive extrapolation lengths, the coefficients $\beta$ and $k$ are compacted into the single factor $\eta_-$, according to \eref{Coef2_eta}. In terms of $\beta$ and $k$ there are three distinct situations: $k<1$, $k=1$ and $k>1$. From the discussion above it directly follows that the energy turns to zero for $k=1$ {\em only} with the additional demand on the cubicity to satisfy the relation
\begin{equation}\label{Equation2}
f\!\left(\frac{\beta}{2}\,,\Lambda\!\right)=0.
\end{equation}
From this equation a critical extrapolation length $\Lambda_0$ at which the zero energy is reached \cite{Olendski2} can be defined:
\begin{equation}\label{LambdaZeroEnergy1}
\Lambda_0\left(\beta\right)=-\frac{1}{\sqrt{2\beta}}\,{\rm dc}\!\!\left(\frac{1}{2}\sqrt{\frac{\beta}{2}},2^{1\left/2\right.}\right)\!{\rm ns}\!\!\left(\frac{1}{2}\sqrt{\frac{\beta}{2}},2^{1\left/2\right.}\right).
\end{equation}
It is a quantitative measure of the exact cancelling by the nonlinear GL term of the temperature increase produced by the negative extrapolation length and, as such, it allows to understand the interaction of the counterbalancing influences of the cubicity and the surface effects. Its limit for the small nonlinearities reads:
\begin{equation}\label{LambdaZeroEnergy3}
\Lambda_{0}\left(\beta\right)=-\frac{2}{\beta},\quad\beta\ll 1.
\end{equation}
This result is understandable since in the linear regime the energy for the large negative de Gennes distances lies only slightly below zero \cite{Olendski3,AlHashimi1} and, accordingly, small effective potential \eref{EffectivePotential1} is sufficient to turn it to zero and subsequently to push into the positive territory. The magnitude of the negative $\Lambda_{0}$ grows together with the cubicity until it reaches zero at
\begin{equation}\label{CriticalBeta1}
\beta_0=4{\bf K}^2\!\left(2^{-1\left/2\right.}\right)=\frac{\Gamma^4\!\left(1/4\right)}{4\pi}=13.7504,
\end{equation}
where $\Gamma(x)$ is $\Gamma$-function \cite{Gradshteyn1,Abramowitz1}. Vanishing $\Lambda_0\left(\beta_0\right)$ means that all critical  lengths with $\beta>\beta_0$, even though mathematically correct, are unphysical ones. In other words, the effective potential \eref{EffectivePotential1} with $\beta=\beta_0$ (and, of course, $k=1$) completely compensates an infinitely large temperature increase produced by the infinitely small negative extrapolation length. Any superconducting state with $\beta>\beta_0$ and $k=1$ has its temperature {\em always} smaller than the critical temperature of the bulk material, 
\begin{equation}\label{Condition1}
T\!\left|_{\beta>\beta_0}<T_c\right.,
\end{equation}
regardless of the surface contribution. From this point of view, the cubicity $\beta$ of the GL equation as compared to the de Gennes boundary is a stronger factor affecting the properties of the superconducting film; however, as it will be shown in the next paragraph, a substantial role in this drastic energy increase is played by the de Gennes distance itself. Critical extrapolation length $\Lambda_0$ is shown in panel (a) of \fref{Fig6}  where the states above (below) the curve correspond to the temperature higher (lower) than $T_c$. 

It might look surprising that the {\em finite} cubicity is able to bring into the area $T<T_c$ the state with any vanishingly small negative $\Lambda$ when, at $\beta=0$, its energy tends to the negative {\em infinity}, according to \eref{Robin1}. For example, a comparison with the other configurations shows that the critical extrapolation length $\Lambda_0(B)$, which is calculated from the linear GL equation for the quantum dot or antidot in the uniform magnetic field $\bf B$, asymptotically tends to zero from below  only  for the intensities $B$ approaching infinity \cite{Olendski2}. To resolve this seeming paradox, one needs to analyse the form of the effective potential \eref{EffectivePotential1} for the small negative de Gennes distance that, in the linear case, $\beta=0$, attracts the order parameter to the edges and, in this way, delivers the larger ability for  the potential to repulse the Cooper pairs from the walls. Thus, the extrapolation length tending to zero from the left produces at nonzero cubicity two contradicting to each other effects: on the one hand, it pushes the energy downwards by localizing the order parameter at the interfaces and, on the other hand, as a consequence of the previous action, it simultaneously provides the effective potential $V_{eff}(x)$, equation \eref{EffectivePotential1}, with the larger ability to repel the function $\chi(x)$ from the film edges with the corresponding increase of the energy. The smaller the magnitude of the negative de Gennes distance is, the larger the negative energy becomes and the stronger localization at the surfaces is but, at the same time, just due to this stronger localization, the faster the effective potential $V_{eff}(x)$ grows with the small increasing $\beta$. In other words, potential \eref{EffectivePotential1}  in its effort to push the energy upwards and the order parameter {\em away} from the edges is actually aided by the small negative length $\Lambda$ that facilitates this energy growth by pulling the order parameter {\em to} the interfaces with the corresponding increase of its amplitude and in this way subjecting the function $\chi(x)$ to the opposite influence of the effective potential. This fine interaction between the cubicity and de Gennes distance results ultimately in the drawing out of the energy from any arbitrary large negative values.

Concerning equation \eref{LambdaZeroEnergy1} and \fref{Fig6}(a), it is necessary to elaborate the following remark already mentioned in passing while discussing \eref{CriticalBeta1}. Since the Jacobi elliptic functions - in particular, ${\rm sn}(u,k)$, ${\rm cn}(u,k)$ and ${\rm dn}(u,k)$, which enter \eref{LambdaZeroEnergy1}, - are doubly periodic functions of its argument $u$, for each $\Lambda_0$ from \fref{Fig6}(a) there exists a countably infinite set of its counterparts with the same value achieved at the larger cubicities $\beta$. However, physically all these critical extrapolation lengths for $\beta>\beta_0$ are spurious ones. This can be seen, for example, from the fact that their corresponding order parameters exhibit infinite discontinuities at some points inside the film, $-1/2\le x\le1/2$. To save space, we do not plot these unphysical functions here presenting instead in panel (b) of \fref{Fig6} an evolution of the zero-energy order parameter 
\begin{equation}\label{SolutionZero2}
\chi_0\left(x\right)={\rm nd}\!\!\left(\!\sqrt{\frac{\beta}{2}}\,x,2^{1\left/2\right.}\!\right)
\end{equation}
with the growing cubicity. It exhibits a smooth transformation from the flat Neumann shape at $\beta=0$, when, according to \eref{LambdaZeroEnergy1} and \eref{LambdaZeroEnergy3}, the critical extrapolation length is equal to the negative infinity, to the linear superposition of the two $\delta$-functions, equation \eref{OrderCritical1}, at $\beta=\beta_0$ when the critical de Gennes distance tends to zero from the left. For any cubicity larger than the critical one, $\beta>\beta_0$, these infinities migrate from the edges into the middle of the film turning in this way the corresponding solutions into the spurious ones. 

Next, consider the cases of $k\ne 1$. When the integration factor $k$ is less than unity, $k<1$, the corresponding solution always exists with its energy lying below zero for all cubicities. For the large nonlinear term, $\beta\rightarrow\infty$, it ceases to depend on the cubicity approaching asymptotically a solution of the equation
\begin{eqnarray}
&&\sqrt{\frac{\pi^2|E|}{1-k^2}}\left(1+k^2\right){\rm sn}\!\!\left(\frac{1}{2}\sqrt{\frac{\pi^2|E|}{1-k^2}},\sqrt{1+k^2}\right)\!{\rm cn}\!\!\left(\frac{1}{2}\sqrt{\frac{\pi^2|E|}{1-k^2}},\sqrt{1+k^2}\right)\nonumber\\
\label{Asymptotics1}
&&+\frac{1}{\Lambda}\,{\rm dn}\!\!\left(\frac{1}{2}\sqrt{\frac{\pi^2|E|}{1-k^2}},\sqrt{1+k^2}\right)=0,\quad 0<k<1,\quad\beta\rightarrow\infty,
\end{eqnarray}
that is derived from \eref{Coef2_eta} - \eref{EigenvalueNegative1}. For the smaller densities the saturation energy (temperature $T$) is smaller (larger). Corresponding order parameter is given then as:
\begin{equation}\label{AsymptoticFunction1}
\fl\chi(x)=\sqrt{\frac{2\pi^2|E|k^2}{\beta\!\left(1-k^2\right)}}\,{\rm nd}\!\!\left(\sqrt{\frac{\pi^2|E|}{1-k^2}}\,x,\sqrt{1+k^2}\right),\quad 0<k<1,\quad\beta\rightarrow\infty,
\end{equation}
what means that the effective potential $V_{eff}(x)$ in this limit becomes $\beta$ independent:
\begin{equation}\label{AsymptoticPotential1}
\fl V_{eff}(x)=\frac{2\pi^2|E|k^2}{1-k^2}\,{\rm nd}^2\!\!\left(\sqrt{\frac{\pi^2|E|}{1-k^2}}\,x,\sqrt{1+k^2}\right),\quad 0<k<1,\quad\beta\rightarrow\infty,
\end{equation}
and it is too weak to subdue the state sufficiently enough for its temperature to lower to $T_c$. On the other hand, for $k>1$ the energy can not grow more than the threshold value $E_{TH}$ defined by the cubicity and density:
\begin{equation}\label{Condition9}
E<E_{TH}\equiv-\frac{1}{2\pi^2}\,\beta(k^2-1),\quad k>1,
\end{equation}
as it directly follows from \eref{Coef2_eta} and \eref{Coef2_zeta}. In fact, the state ceases to exist at the energies smaller than $E_{TH}$ since \eref{EigenvalueNegative1} in this case contains the Jacobi functions of the form
$$
{\rm sn}\!\left(\sqrt{\frac{1}{E-E_{TH}}},\eta_-\right)
$$
that, similar to their trigonometric counterparts \cite{Fikhtengolts1}, possess infnitely many oscillations for $E\rightarrow E_{TH}$. Upon its growth with the cubicity, the eigenvalue of \eref{EigenvalueNegative1} approaches one of the zeros of these oscillations and merges with it after which the two corresponding {\em real} energies transform into the two {\em complex} conjugate values. Since the present research deals with the {\em real} energies and order parameters only, this merger is considered as a disappearance of the corresponding state. Physically, a destruction of the superconductivity for these parameters is explained by the fact that the decreasing temperature $T$ can not support the fixed large density $n_s$ of the superconducting carriers with its concentration near the film interfaces.
\begin{figure}
\centering
\includegraphics[width=0.99\columnwidth]{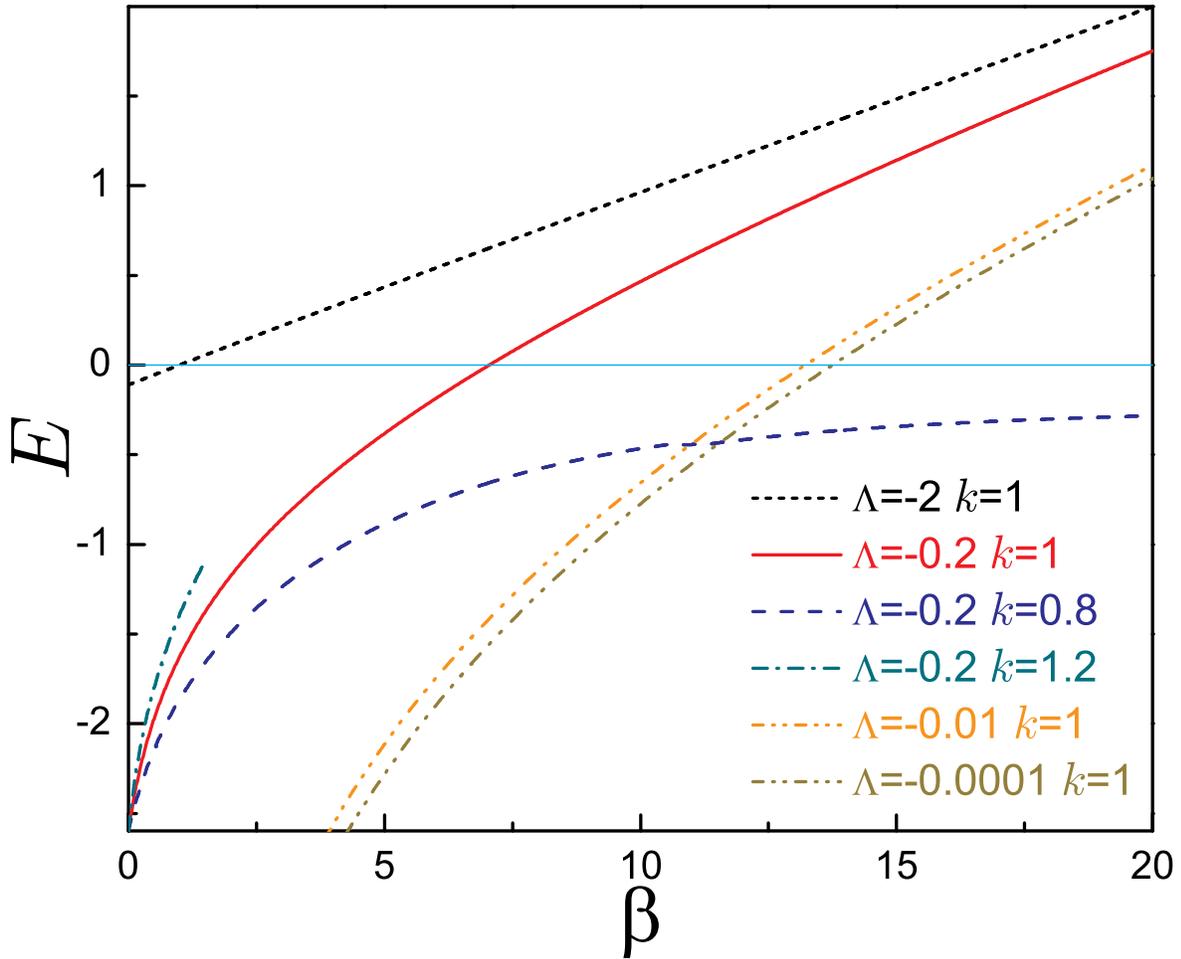}
\caption{\label{Fig7}
Energy $E$ as a function of the cubicity $\beta$ for several values of the density and negative extrapolation length where the dotted curve is for $\Lambda=-2$ and $k=1$, the solid line is for $\Lambda=-0.2$ and $k=1$, the dashed line is for $\Lambda=-0.2$ and $k=0.8$, the dash-dotted line is for $\Lambda=-0.2$ and $k=1.2$, and two dash-dot-dotted lines are for $k=1$ and $\Lambda=-0.01$ (upper curve) or $\Lambda=-0.0001$ (lower curve). Thin horizontal line denotes zero energy.
}
\end{figure}

\begin{figure}
\centering
\includegraphics[width=0.99\columnwidth]{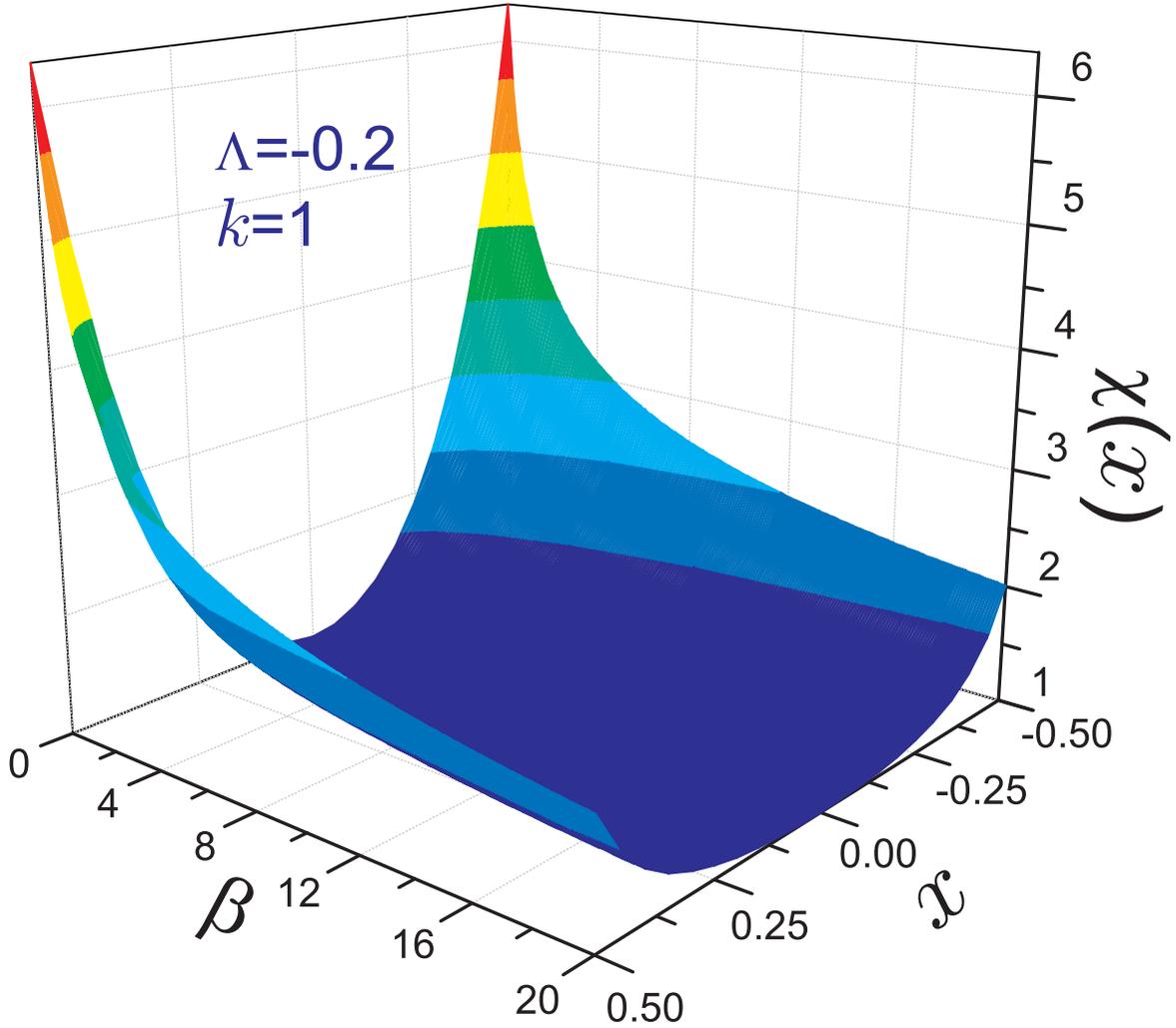}
\caption{\label{Fig8}
Order parameter $\chi(x)$ as a function of the cubicity $\beta$ for $\Lambda=-0.2$ and $k=1$. Corresponding energy is depicted by the solid line in \fref{Fig7}.
}
\end{figure}

To exemplify the results obtained above, \fref{Fig7} depicts energy as a function of the cubicity for several miscellaneous combinations of the negative extrapolation lengths and densities. It is seen that for $k<1$ (dashed line) the energy saturates with the increasing $\beta$ to its asymptotic value calculated from \eref{Asymptotics1} while for $k>1$ (dash-dotted curve) the state ceases to exist on its energy approach to $E_{TH}$. A comparison of the curves with $k=1$ and different magnitudes of the negative de Gennes distances (solid, dotted and two dash-dot-dotted lines) vividly manifests a convergence of the zero-energy cubicity to its critical value $\beta_0$ from \eref{CriticalBeta1}; namely, for the large enough negative $\Lambda$ only the small cubicity is needed to bring the energy to zero (dotted line in \fref{Fig7}) while for the very small magnitudes of the de Gennes distance, for example, depicted by the two dash-dot-dotted lines, the energies of the corresponding states that differ by four orders of magnitudes in the liear regime, $\beta=0$, cross zero at about the same $\beta$ being very close to $\beta_0$. After passing zero, the energy continues to rise with its order parameter given by
\begin{equation}\label{Chi1}
\left.\chi(x)\right|_{k=1}={\rm dc}\!\!\left(\!\sqrt{\frac{\beta}{2}}\,x,\sqrt{\frac{2\pi^2E-\beta}{\beta}}\,\right).
\end{equation}
As it follows from \eref{Limit5}, for the very large nonlinearities, it diverges to the infinity with $\beta/\pi^2$ being its principal term, and the order parameter tends to unity, equation \eref{Chi3}. Thus, since the concentration with $k=1$ for the large cubicities corresponds to the boundary-free situation, the state with only this density  is allowed, for the negative de Gennes distances, to cross the bulk critical temperature with the increasing $\beta$. Evolution with the cubicity of the order parameter $\chi(x)$ corresponding to the solid line of \fref{Fig7} is shown in \fref{Fig8}. It is seen that its upward concavity decreases with the growing parameter $\beta$ until, at the very large nonlinearities, it becomes, according to \eref{Chi3}, completely flat.

Finally, we provide a very instructive comparison of the above results for the temperature $T$ passing through the bulk critical temperature $T_c$ with the predictions derived from \eref{ChiEq1} where the energy $E$ is explicitly set to zero:
\begin{equation}\label{ChiEq3}
\chi''\left(x\right)-\beta\chi^3\left(x\right)=0.
\end{equation}
Its corresponding eigenfunction $\chi^{(0)}(x)$ reads:
\begin{equation}\label{Solution3}
\chi^{(0)}(x)=k\,{\rm nd}\!\left(k\sqrt{\frac{\beta}{2}}\,x,2^{1\left/2\right.}\!\right).
\end{equation}
Then, it directly follows from the boundary condition 
\begin{equation}\label{Equation3}
f\!\left(\frac{k^2\beta}{2}\,,\Lambda\!\right)=0
\end{equation}
that the corresponding critical extrapolation length $\Lambda^{(0)}$ is:
\begin{equation}\label{LambdaZeroEnergy2}
\Lambda^{(0)}\left(k,\beta\right)=-\frac{1}{k\sqrt{2\beta}}\,{\rm dc}\!\left(\frac{1}{2}\,k\sqrt{\frac{\beta}{2}},2^{1\left/2\right.}\!\right){\rm ns}\!\left(\frac{1}{2}\,k\sqrt{\frac{\beta}{2}},2^{1\left/2\right.}\!\right).
\end{equation}
A comparison with the previous derivation shows the difference between the two results; namely, while \eref{LambdaZeroEnergy1} is obtained under the strict condition that the integration factor $k$ is unity and cubicity $\beta$ obeys \eref{Equation2},  the claim of \eref{LambdaZeroEnergy2} is that the zero energy can be achieved at any arbitrary values of the positive parameters $\beta$ and $k$ provided that their product $k^2\beta$ satisfies \eref{Equation3}. It is easy to understand the origin of this wrong conclusion; namely, when the energy is bluntly put to zero in \eref{ChiEq1}, resulting equation \eref{ChiEq3} and its solution~\eref{Solution3} know nothing about the quasi-peculiarity in \eref{Coef2_zeta} and its elimination by the ansatz from \eref{FunctionE}. Equations~\eref{SolutionNegative1} - \eref{EigenvalueNegative1}, \eref{Equation2}, \eref{LambdaZeroEnergy1} and \eref{SolutionZero2} for the negative $E$ contain much more information about the temperature behaviour close to $T_c$ than their zero-energy counterparts \eref{Solution3} - \eref{LambdaZeroEnergy2}. Due to their limited knowledge, the latter equations  erroneously advice us that the zero energy can be achieved at any magnitudes of $\beta$ and $k$ if the product $k^2\beta$ obeys \eref{Equation3} while in the reality the film temperature is equal to the bulk critical temperature $T_c$ for the unity factor $k$ only with the cubicity $\beta$ satisfying \eref{Equation2}.

\section{Concluding remarks}\label{sec_Conclusions}
Exact solutions of the GL equation for the superconducting film with its edges subjected to the Robin-type boundary condition \eref{BoundaryCondition2} revealed strong dependence of the temperature $T$ on the interaction between the de Gennes distance and cubicity with, in addition, very substantial influence of the superconducting carriers density $n_s$. Depending on these parameters, a rich spectrum of the temperature $T$ is obtained and analysed in detail. For example, the interplay of the negative extrapolation length and nonlinearity causes the decreasing with the growing cubicity temperature $T$ to cross its bulk critical value at the unique carrier concentration {\em only} while for the smaller densities its stays above $T_c$ for all cubicites, and the superconductivity is destroyed by the increasing coefficient $\beta$ for the larger Cooper pairs concentration. The uniqueness of this density stems from the fact that only for it the increasing cubicity transforms the solution of the GL equation for the film into the one corresponding to the bulk superconductor.

There are several possible ways of varying the de Gennes distance and magnitude of the nonlinear GL term. Experiment suggests that  the value of the extrapolation length can be controlled by the sample preparation procedure \cite{Kozhevnikov1}. Even more promising from this point of view is its manipulation by the applied gate voltage, according to \eref{TotalExtrapolationLength1} and \eref{PotentialU}. Miscellaneous aspects of the influence of the   electrostatic fields on the superconducting films have been the subject of the experimental investigations for more than fifty years \cite{Glover1,Fiory1,Mannhart1,Xi1,Matijasevic1,Ahn1,Dhoot1}; in particular, the change of the critical temperature under nonzero $\mbox{\boldmath${\cal E}$}$ has been observed  \cite{Glover1,Matijasevic1,Dhoot1}. However, despite its long history of research, electric field effects on superconductors are still far from being understood \cite{Chandrasekhar1,Aligia1,Frey1,Hirsch1,Onuki1,Kolacek1}. With regards to the cubicity $\beta$, the methods of its variation follow straightforwardly from definition \eref{CoeffBeta1} for the pure metals \cite{Gorkov1,Schmidt1} and, especially, its alloy modification \cite{deGennes1,Schmidt1,Gorkov2}; namely, different impurity doping drastically changes the mean free path $l$ that becomes much smaller than the coherence length $\xi(0)$ thus leading to the substantial increase of the nonlinear influence. In addition, the range of validity of the GL theory is much wider for the 'dirty' superconductors \cite{Schmidt1}. Accordingly, it is believed that applying the gate voltage to the films composed with the different concentration of impurities, one can experimentally observe the predicted phenomena.

\end{document}